\documentclass[twocolumn,floatfix,showpacs,prb,superscriptaddress,eqsecnum]{revtex4}
\usepackage{graphicx}
\usepackage{amsmath}
\usepackage{amssymb}
\usepackage{bm}

\usepackage{color}

\begin{document}

\title{Quantum Hall effect in a one-dimensional dynamical system}
\author{J. P. Dahlhaus}
\affiliation{Instituut-Lorentz, Universiteit Leiden, P.O. Box 9506, 2300 RA Leiden, The Netherlands}
\author{J. M. Edge}
\affiliation{Instituut-Lorentz, Universiteit Leiden, P.O. Box 9506, 2300 RA Leiden, The Netherlands}
\author{J. Tworzyd\l o}
\affiliation{Institute of Theoretical Physics, University of Warsaw, Ho\.{z}a 69, 00--681 Warsaw, Poland}
\author{ C. W. J. Beenakker}
\affiliation{Instituut-Lorentz, Universiteit Leiden, P.O. Box 9506, 2300 RA Leiden, The Netherlands}
\date{June 2011}
\begin{abstract}
We construct a periodically time-dependent Hamiltonian with a phase transition in the quantum Hall universality class. One spatial dimension can be eliminated by introducing a second incommensurate driving frequency, so that we can study the quantum Hall effect in a one-dimensional (1D) system. This reduction to 1D is very efficient computationally and would make it possible to perform experiments on the 2D quantum Hall effect using cold atoms in a 1D optical lattice.\end{abstract}
\maketitle

\section{Introduction}
\label{sec:intro}

The disorder-induced localization-delocalization transition in the quantum Hall effect is the oldest and best-known example of a topological phase transition.\cite{Kli80,Huc95} The transition is called topological because it is associated with a  change in a topological invariant, the Chern number, which counts the number of edge states and the quantized value of the Hall conductance.\cite{Tho82} Since there is still no analytical theory for the quantum Hall phase transition, computer simulations are needed to calculate the scaling law and critical exponent associated with the diverging localization length at the transition. The two-dimensional (2D) network model of Chalker and Coddington has been the primary tool for these studies for more than two decades.\cite{Cha88,Kra05,Slevin2009}

In this paper we introduce an alternative \textit{one-dimensional} (1D) model of the quantum Hall phase transition. The model is stroboscopic, with a Hamiltonian that is driven quasiperiodically with two incommensurate driving frequencies. It is a variation on the quantum kicked rotator,\cite{Cas79,Fishman1982,Shepelyansky1983} used to study the 3D Anderson metal-insulator transition of atomic matter waves in a 1D optical lattice.\cite{Cha08,Lemarie2010,Sad08} Stroboscopic models of quantum phase transitions have received much attention recently,\cite{Kit10a,Kit10b,Ino10,Lindner2011,Jia11,Obuse2011} but the dimensional reduction considered here was not yet explored.

Usually the quantum Hall effect is due to the quantization of cyclotron orbits in Landau levels. It is possible to simulate a Lorentz force acting on neutral atoms in a 2D optical lattice,\cite{Coo08,Lin09,Dal10} but in a 1D lattice we need a quantum Hall effect without Landau levels.\cite{Hal88} This socalled quantum \textit{anomalous} Hall effect appears in the Qi-Wu-Zhang (QWZ) model of a spin-$1/2$ coupled to orbit and to a uniform magnetization. While the topological invariant in this model takes on only the three values $0,\pm 1$, the phase transitions are in the same $\mathbb{Z}$ universality class as the usual quantum Hall effect.

In the next two sections we formulate the stroboscopic model of the quantum Hall effect, first in 2D (Sec.\ \ref{sec:model}) and then reduced to 1D (Sec.\ \ref{sec:mapping-1d-model}). We obtain the model by starting from the QWZ Hamiltonian, but we also show how it is related to the quantum kicked rotator (upon exchange of position and momentum). 

In Sec.\ \ref{dyn_sim_of_kicked_rot} we perform numerical simulations of the spreading of a 1D wave packet to identify the localization-delocalization transitions. While the translationally invariant QWZ model has three quantum Hall transitions, we find four transitions because one is split by disorder. We verify one-parameter scaling of the time-dependent diffusion coefficient and calculate the critical exponent. The result is consistent with the most accurate value obtained from the Chalker-Coddington model.\cite{Slevin2009} 

To further support that these are topological phase transitions in the quantum Hall universality class we calculate the Hall conductance as well as the ${\mathbb Z}$ topological invariant in Sec.\ \ref{Topology}. We conclude by discussing the possibilities for the realization of the quantum Hall effect in a 1D optical lattice.

\section{Formulation of the 2D stroboscopic model}
\label{sec:model}

\subsection{Quantum anomalous Hall effect}

In this subsection we summarize the QWZ model\cite{Qi06} of the quantum anomalous Hall effect, on which we base the stroboscopic model described in the next subsection.

The QWZ model describes two spin bands of a magnetic insulator on a two-dimensional (2D) square lattice. The crystal momentum $\bm{p}=(p_{1},p_{2})$ varies over the Brillouin zone $-\pi\hbar/a<p_{1},p_{2}<\pi\hbar/a$. The Hamiltonian has the form $\bm{u}\cdot\bm{\sigma}$, with $\bm{\sigma}=(\sigma_{x},\sigma_{y},\sigma_{z})$ a vector of Pauli matrices and 
\begin{equation}
\bm{u}(\bm{p})  =  K \bigl(   \sin p_1,    \sin p_2,     \beta[\mu-\cos p_1-\cos p_2]    \bigr)\label{updef}
\end{equation}
a momentum-dependent vector that couples the spin bands. (We have set $\hbar$ and $a$ both equal to unity.) The dispersion relation is $E_{\pm}(\bm{p})=\pm u(\bm{p})$, with $u=|\bm{u}|$ the norm of the vector $\bm{u}$. We fix the Fermi level at zero, in the middle of the energy gap.

Eq.\ \eqref{updef} contains three parameters, $K,\beta,\mu$. The parameter $K$ sets the strength of the spin-orbit coupling. Time-reversal symmetry is broken by a nonzero $\beta$, representing a magnetization perpendicular to the 2D plane. The Hall conductance $G_{\rm H}$ is quantized at\cite{Qi06}
\begin{equation}
G_{\rm H}=\frac{e^{2}}{h}\times\left\{\begin{array}{cl}
\mbox{}{\rm sign}\,\beta\mu&{\rm if}\;\;|\mu|<2,\\
0&{\rm if}\;\;|\mu|>2.
\end{array}\right.\label{QWZsigmaxy}
\end{equation}
This quantum Hall effect is called ``anomalous'', because it does not originate from Landau level quantization.

The value of $G_{\rm H}$ is a topological invariant,\cite{Qi06} meaning that it is insensitive to variations of the Hamiltonian that do not close the energy gap. Since the gap can only close if $u(\bm{p})$ vanishes for some $\bm{p}$, the Hamiltonian
\begin{equation}
H_{0}(\bm{p})={\cal T}(u)\,\bm{u}\cdot\bm{\sigma}\label{H0def}
\end{equation}
has the same quantized Hall conductance \eqref{QWZsigmaxy} if the function ${\cal T}(u)$ is positive definite. We will make use of this freedom in order to flatten the spin bands, by choosing a function ${\cal T}(u)$ which decays for large $u$.

The Hamiltonian $H_{0}(\bm{p})$ describes a clean system. The effects of electrostatic disorder are included by adding the scalar potential $V(\bm{x})$. The 2D coordinate $\bm{x}=(x_{1},x_{2})$ is measured in units of $a$, while momentum $\bm{p}=(p_{1},p_{2})$ is measured in units of $\hbar/a$, so their commutator is $[x_{n},p_{m}]=i\delta_{nm}$.

\subsection{Stroboscopic Hamiltonian}

This completes the description of the QWZ model. We now introduce a periodic time dependence by multiplying $H_{0}$ with the stroboscopic function $\tau\sum_{n}\delta(t-n\tau)$, while keeping the scalar potential time-independent. We thus arrive at the stroboscopic Hamiltonian
\begin{equation}
{\cal H}(t)=V(\bm{x})+H_{0}(\bm{p})\sum_{n=-\infty}^{\infty}\delta(t-n),\label{Htdef}
\end{equation}
where we have set the period $\tau$ equal to unity.

For the choice of ${\cal T}(u)$ and $V(\bm{x})$ we are guided by the tight-binding representation given in App.\ \ref{sec:HoppingModel}. We use
\begin{equation}
{\cal T}(u)=\frac{2\arctan u}{u},
\label{eq:Tarctan}
\end{equation}
which has a tight-binding representation with nearest-neighbor hopping. For the scalar potential $V(\bm{x})$ we take a separable form,
\begin{equation}
V(\bm{x})=\sum_{i=1}^{2} V_{i}(x_{i}),\label{Vxdef}
\end{equation}
with $V_{i}(x_{i})$ a low-order polynomial in $x_{i}$. Such a simple potential produces quasi-random on-site disorder in the tight-binding representation.

\subsection{Relation to quantum kicked rotator}
\label{sec:QKR}

The quantum kicked rotator is a particle moving freely along a circle, with moment of inertia $I$, being subjected periodically (with period $\tau$) to a kick whose strength depends $\propto\cos\theta$ on the angular coordinate $\theta$. The quantum mechanical Hamiltonian is\cite{Cas79,Izr90}
\begin{equation}
H(t)=-\frac{\hbar^{2}}{2I}\frac{\partial^{2}}{\partial \theta^{2}}+\frac{KI}{\tau}\cos\theta\sum_{n=-\infty}^{\infty}\delta(t-n\tau).\label{HQKR}
\end{equation}
The stroboscopic Hamiltonian \eqref{Htdef} has the same general form, upon substitution of $\theta\mapsto\bm{p}$, with the extension from 1D to 2D and with the addition of a spin degree of freedom in the kicking term.

A 1D spinfull  kicked rotator has been used to study the effects of spin-orbit coupling on quantum localization. \cite{Scharf1989,Tha94,Mas94,Oss04,Bar05,Bar07} (Because in the kicked rotator the variable which localizes is momentum rather than position, one speaks of \textit{dynamical} localization.) In these 1D studies there was only a topologically trivial phase, while --- as we shall see --- the present 2D model exhibits a topological phase transition.

\subsection{Floquet operator}
\label{sec:Floquet}

The evolution $\Psi(t+1)={\mathcal F}\Psi(t)$ of the wave function $\Psi(t)$ over one period is described by the Floquet operator $\mathcal{F}$. Integration of the Schr\"{o}dinger equation $i\partial\Psi/\partial t={\cal H}(t)\Psi(t)$ gives the Floquet operator as the product
\begin{equation}
{\cal F}=e^{-iH_{0}(\bm{p})}e^{-iV(i\partial_{\bm{p}})},\label{floquet_as_prod}
\end{equation}
with $i\partial_{\bm{p}}\equiv i\partial/\partial\bm{p}$ the position operator $\bm{x}$ in momentum representation.

\begin{figure}[tb]
\centerline{\includegraphics[width=0.8\linewidth]{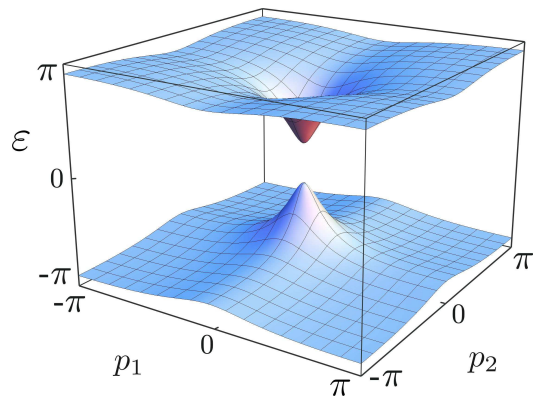}}
\caption{\label{fig:VectorField}
Momentum dependence of the quasi-energy \eqref{epsV0} for zero disorder potential, calculated from Eqs.\ \eqref{updef} and \eqref{eq:Tarctan} for $K=2$, $\beta=0.8$, $\mu=1.9$. At the center of the Brillouin zone the Dirac cone emerges, which will be fully formed when the gap closes at $\mu=2$.
}
\end{figure}

The eigenvalues of the unitary operator ${\cal F}$ are phase factors $e^{-i\varepsilon}$. The phase shift $\varepsilon\in[-\pi,\pi)$ plays the role of energy (in units of $\hbar/\tau$), and is therefore called a quasi-energy. For $V\equiv 0$ the quasi-energy is an eigenvalue of $H_{0}$, hence
\begin{equation}
\varepsilon=\pm u{\cal T}(u),\;\;{\rm for}\;\;V\equiv 0.\label{epsV0}
\end{equation}
The $\bm{p}$-dependence of the two bands is plotted in Fig.\ \ref{fig:VectorField}. The emerging Dirac cone is clearly visible. Away from the cone the bands are quite flat, which is a convenient feature of our choice \eqref{eq:Tarctan} of ${\cal T}(u)$.

More generally, for nonzero $V$, the $2\pi$-periodicity of $H_{0}(\bm{p})$ implies that the eigenstates
\begin{equation}
\Psi_{\bm{q}}(\bm{p})=e^{-i\bm{p}\cdot\bm{q}}\chi_{\bm{q}}(\bm{p})\label{psichi}
\end{equation}
of ${\cal F}$ are labeled by a Bloch vector $\bm{q}$ in the Brillouin zone $-\pi<q_{1},q_{2}<\pi$. The function $\chi_{\bm{q}}(\bm{p})$ is a $2\pi$-periodic eigenstate of
\begin{equation}
{\cal F}_{\bm{q}}=e^{-iH_{0}(\bm{p})}e^{-iV(i\partial_{\bm{p}}+\bm{q})}.\label{floquet_q}
\end{equation}
A convenient basis for the functions $\chi_{\bm{q}}(\bm{p})$ is formed by the eigenfunctions $\exp(-i\bm{m}\cdot\bm{p})$ of $\bm{x}$. The $2\pi$-periodicity of $\chi_{\bm{q}}$ requires that the vector $\bm{m}=(m_{1},m_{2})$ contains integers.

\section{Mapping onto a 1D model}
\label{sec:mapping-1d-model}

The quantum kicked rotator in $d$ dimensions can be simulated in one single dimension by means of $d$ incommensurate driving frequencies.\cite{Shepelyansky1983,Casati1989} We apply this dimensional reduction to our stroboscopic model of the quantum Hall effect.

We take a linear potential in the variable $x_{2}$,
\begin{equation}
V(\bm{x})=V_{1}(x_{1})-\omega x_{2},\label{VV1def}
\end{equation}
with $\omega/2\pi$ an irrational number $\in(0,1)$. During one period the momentum $p_2$ is incremented to $p_2+\omega $ (modulo $2\pi$), so $\omega$ is an incommensurate driving frequency. An initial state
\begin{equation}
\Psi(p_1,p_2,t=0)=\psi(p_1,t=0)\delta(p_2-\alpha).\label{chit0}
\end{equation}
evolves as
\begin{align}
\Psi(p_{1},p_{2},t)={}&e^{-iH_{0}(p_{1},\omega t+\alpha)}e^{-iV_{1}(x_{1})}\psi(p_{1},t-1)\nonumber\\
&\times\delta(p_{2}-\omega t-\alpha).\label{psiomegat}
\end{align}
We may therefore replace the 2D dynamics by a 1D dynamics with a time-dependent Floquet operator
\begin{align}
&{\cal F}(t)=e^{-iH_{0}(p_{1},\omega t+\alpha)}e^{-iV_{1}(i\partial_{p_{1}})},\label{Floquett1}\\
&{\cal F}_{q}(t)=e^{-iH_{0}(p_{1},\omega t+\alpha)}e^{-iV_{1}(i\partial_{p_{1}}+q)}.\label{Floquett2}
\end{align}
This reduction from two dimensions to one dimension greatly simplifies the numerical simulation of the quantum Hall effect.

For the potential in the remaining dimension we take a quadratic form,
\begin{equation}
V_{1}(x_{1})=\tfrac{1}{2}\lambda (x_{1}-x_{0})^{2},\label{V1def}
\end{equation}
with $x_{0}$ an arbitrary offset and $\lambda,\omega,2\pi$ an incommensurate triplet. (We take $\lambda=2$, $\omega=2\pi/\sqrt{5}$.) From studies of the $d$-dimensional quantum kicked rotator it is known that such a simple potential, which is linear in $d-1$ dimensions and nonlinear in one single dimension, provides sufficient randomness for localization.\cite{Bor97}

\section{Localization in the quantum Hall effect}
\label{dyn_sim_of_kicked_rot}

\subsection{Numerical simulation}
\label{numerics}

We base our numerical simulation on the 1D stroboscopic model with two incommensurate frequencies of Sec.\ \ref{sec:mapping-1d-model}. We introduce a Bloch number $q$ and seek the time dependence of the state $\psi(p_{1},t)=e^{-iqp_{1}}\chi_{q}(p_{1},t)$. The state $\chi_{q}(p_{1},t)$ is a $2\pi$-periodic function of $p_{1}$, so it is a superposition of a discrete set of eigenstates $e^{-imp_{1}}$ of $x_{1}$. For numerical purposes this infinite set is truncated to $M$ states, $m\in\{1,2,\ldots M\}$, with periodic boundary conditions at the end points. 

Fourier transformation from eigenstates of $x_{1}$, with eigenvalue $m$, to eigenstates of $p_{1}$, with eigenvalue $2\pi n/M$, amounts to multiplication with the unitary matrix
\begin{equation}
U_{nm}=M^{-1/2}e^{2\pi inm/M},\;\;n,m\in\{1,2,\ldots M\}.\label{Unmdef}
\end{equation}
Calculation of the state $\chi_{q}(x_{1},t)$, for $t$ an integer multiple of $\tau\equiv 1$, requires $2t$ Fourier transformations,
\begin{align}
&\chi_{q}(x_{1},t)=\left(\prod_{t'=0}^{t-1}{\cal F}_{q}(t')\right)\chi_{q}(x_{1},0),\label{chiqtchiq0}\\
&\left[{\cal F}_{q}(t)\right]_{nm}=\sum_{k=1}^{M}U^{\ast}_{kn}e^{-iH_{0}(2\pi k/M,\omega t+\alpha)}U_{km}e^{-iV_{1}(m+q)}.\label{Ftdef}
\end{align}
These operations can be carried out with high efficiency using the fast-Fourier-transform algorithm.\cite{Ket99}

As initial state we choose
\begin{align}
\chi_{q}(x_{1},0)={}&\delta_{x_{0},x_{1}}\left[e^{-i\phi_{0}/2}\cos(\theta_{0}/2)\begin{pmatrix}
0\\
1
\end{pmatrix}\right.\\
&+\left.e^{i\phi_{0}/2}\sin(\theta_{0}/2)\begin{pmatrix}
1\\
0
\end{pmatrix}\right],
\label{chiqt0}
\end{align}
spatially localized at $x_{1}=x_{0}=M/2$ (for even $M$). The angles $\phi_{0},\theta_{0}$ of the initial spin direction are chosen randomly on the unit sphere. 

\subsection{Localization-delocalization transition}
\label{transition}

To search for localization we calculate the expectation value
\begin{equation}
\bigl\langle \bigl(x_{1}(t)-x_{0}\bigr)^{2}\bigr\rangle=\sum_{m=1}^{M}(m-M/2)^{2}|\chi_{q}(x_{1}=m,t)|^{2},\label{x1squared}
\end{equation}
and obtain the mean squared displacement
\begin{equation}
\Delta^{2}(t)=\overline{\bigl\langle \bigl(x_{1}(t)-x_{0}\bigr)^{2}\bigr\rangle}\label{Deltasquared}
\end{equation}
by averaging over some $10^{2}$--$10^{3}$ values of the random parameters $\alpha,q\in\{0,1\}$. We fix $K=2$, $\beta=0.8$ and vary the parameter $\mu$.

The system is localized if  the time-dependent diffusion coefficient
\begin{equation}
D(t)= \frac{\Delta^{2}(t)}{t}\label{Dtdef}
\end{equation}
vanishes in the large-$t$ limit. Delocalization with diffusive propagation corresponds to a non-zero large-time limit of $D(t)$. The quantum Hall phase transition is a localization-delocalization transition, so we would expect a peak in $D(t)$ as a function of $\mu$ at the critical points $\mu_{c}$ where the topological invariant switches from one value to another. In a clean system these values are $\mu_{c}=0,\pm 2$, see Eq.\ \eqref{QWZsigmaxy}.

\begin{figure}[tb]
\centerline{\includegraphics[width=0.9\linewidth]{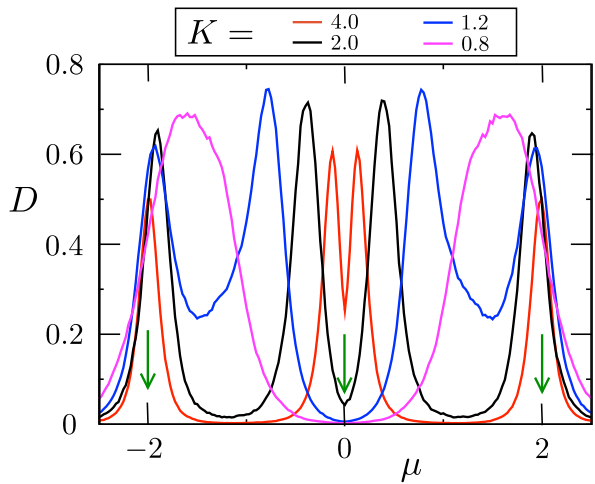}}
\caption{\label{fig:peakmove}
Time-dependent diffusion coefficient \eqref{Dtdef} at $t=3000$ as a function of $\mu$ for $M=1024$. The different values of $K$ range from weak disorder ($K=4$) to strong disorder ($K=0.8$). The peaks signal a localization-delocalization transition. Compared to the three quantum Hall transitions in a clean system (indicated by arrows), the two outer transitions are displaced inwards by disorder, while the central transition is split. The splitting of the two central peaks becomes larger and larger with increasing disorder, until they merge with the outer peaks.
}
\end{figure}

\begin{figure}[tb]
\centerline{\includegraphics[width=0.9\linewidth]{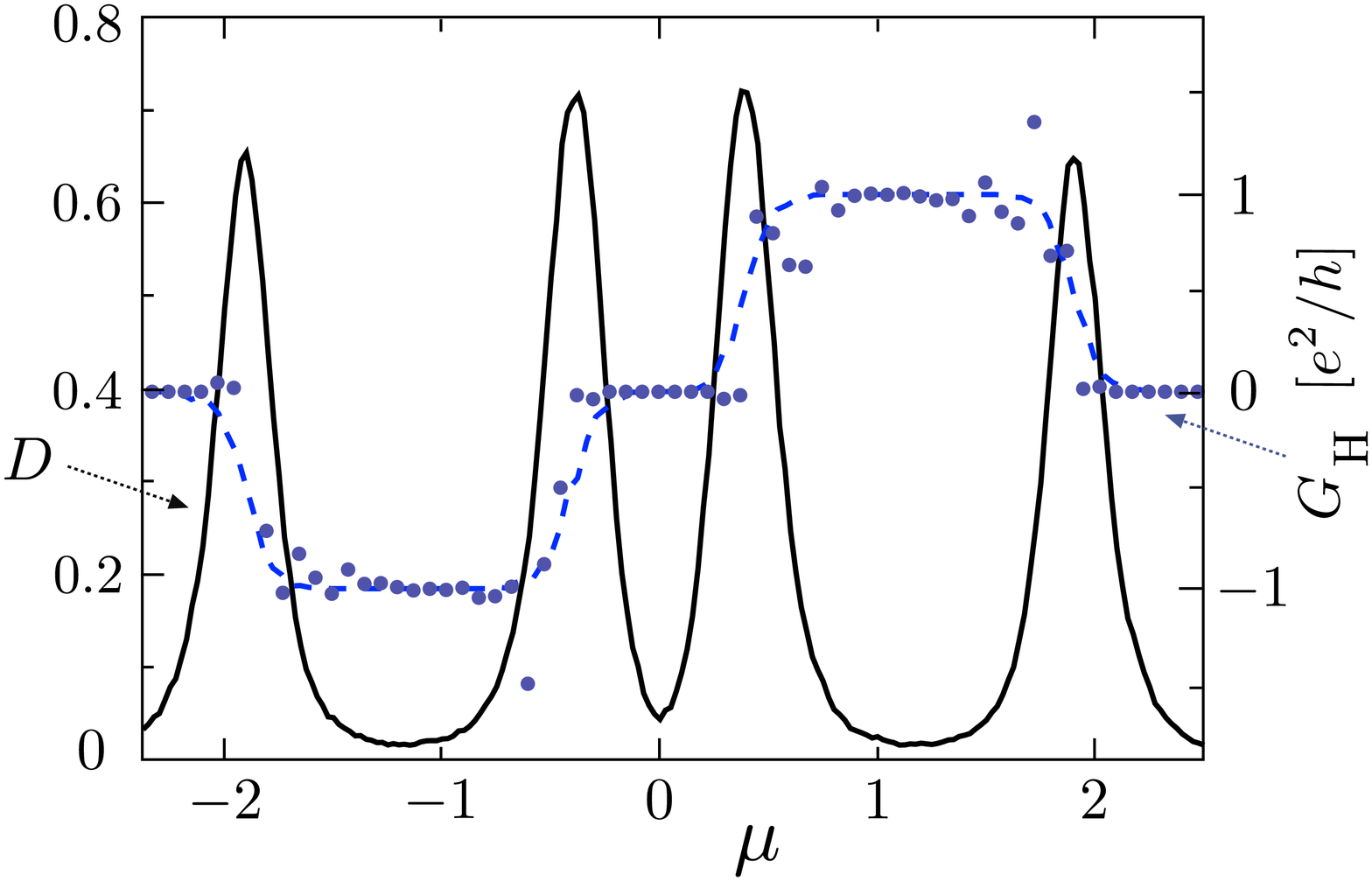}}
\caption{\label{fig:peakscan} 
Left axis: Time-dependent diffusion coefficient for $K=2$ (solid curve, same data as in Fig.\ \ref{fig:peakmove}), showing the four localization-delocalization transitions. Right axis: Four-terminal Hall conductance $G_{\rm H}$ (data points) and topological invariant ${\cal I}$ (dashed curve), calculated in Sec.~\ref{Topology}, to demonstrate that these are \textit{topological} phase transitions.
}
\end{figure}

The data in Fig.\ \ref{fig:peakmove} shows that disorder has two effects: It shifts the outer transitions inwards and splits the central transition, resulting in a total of four peaks. We will demonstrate in Sec.\ \ref{Topology} that these are \textit{topological} phase transitions, by calculating the topological invariant --- which as we can see in Fig.\ \ref{fig:peakscan} switches at each of the transitions.

\subsection{Scaling and critical exponent}
\label{parameter}

In the single-parameter scaling theory of localization all microscopic parameters enter only through a single length scale $\xi$ (the localization length) and the associated energy scale $\delta_{\xi}=(\xi^{d}\rho_{c})^{-1}$ (being the mean level spacing in a $d$-dimensional box of size $\xi$, obtained from the density of states $\rho_{c}$ at the critical energy).\cite{Cha90,Bra96,Huc98} The corresponding scaling law for dynamical localization has the form\cite{Lemarie2009}
\begin{equation}
D(t)= \xi^{2-d}F(\xi^{-d}t),\label{Dtscaling}
\end{equation}
in the large-time limit near the critical point $\mu_{c}$. The localization length $\xi$ diverges as a power law with critical exponent $\nu$ on approaching the transition,
\begin{align}
\xi\propto |\mu-\mu_{c}|^{-\nu}.
\label{eq:CriticalExponent}
\end{align}

The limiting behavior of the function $F(z)$ is $F(z)\propto 1/z$ for $z\rightarrow\infty$ and $F(z)\propto z^{2/d-1}$ for $z\rightarrow 0$. The first limit ensures that the mean squared displacement $\Delta^{2}=tD(t)\rightarrow\xi^{2}$ becomes time independent in the limit $t\rightarrow\infty$ at fixed $\mu-\mu_{c}$. The second limit ensures that, if we send $\mu\rightarrow \mu_{c}$ at fixed $t$, the diffusion coefficient $D(t)\rightarrow t^{2/d-1}$ tends to a finite value. For $d=2$, this value is also time independent, which implies regular diffusion ($D={\rm constant}$) at criticality in two dimensions.

\begin{figure}[tb]
\centerline{\includegraphics[width=0.7\linewidth]{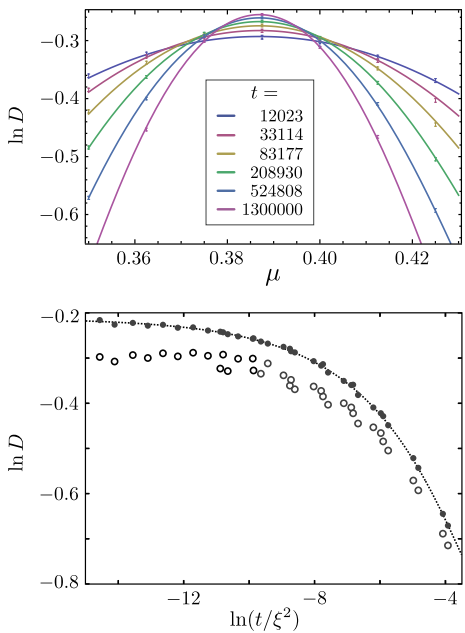}}
\caption{\label{fig_scalingII}
Top panel: Time-dependent diffusion coefficient for different times as a function of $\mu$. The curves are a least-squares fit, used to extract the localization length $\xi(\mu)$ and the critical exponent (see App.\ \ref{finitetime}). In the lower panel time is rescaled, to test the scaling form $D(t,\mu)=F(t/\xi^{2})$ [see Eq.\ \eqref{Dtscaling}]. The open data points do not fully collapse onto a single scaling curve, due to finite-time corrections to scaling. The filled data points include the leading-order correction (see App.\ \ref{finitetime}).
}
\end{figure}

We have performed a finite-time scaling analysis of $D(t)$, similar to Refs.\ \onlinecite{Slevin2009,Lemarie2009}, to obtain the localization length $\xi$ and extract the value of the critical exponent. (See App.\ \ref{finitetime} for details.) We considered times up to $t=1.3\cdot 10^{6}$ for system size $M=2^{13}=8192$. In Fig.\ \ref{fig_scalingII} we show both the unscaled and the scaled data. For the two independent transitions we find $\nu=2.576\pm0.03$ at $\mu_{c}=0.387$ and $\nu=2.565\pm0.03$ at $\mu_{c}=1.903$. Both results agree with $\nu_{\rm QHE}=2.593$, the critical exponent for the quantum Hall phase transition.\cite{Slevin2009}

\section{Hall conductance and topological invariant}
\label{Topology}

The anomalous quantum Hall effect in the absence of disorder ($V\equiv 0$) is characterized by the topological invariant\cite{Qi06}
\begin{equation}
\mathcal{I}=-\frac{1}{4\pi}\int_{-\pi}^{\pi} d p_1 \int_{-\pi}^{\pi} d p_2 \left[\frac{\partial {\bm{\hat{u}}}(\bm{ p})}{\partial  p_1}\times \frac{\partial {\bm{\hat{u}}}(\bm{ p})}{\partial  p_2}\right]\cdot {\bm{\hat{u}}}(\bm{ p}),
\label{eq:Invariantclean}
\end{equation}
with $\bm{\hat{u}}=\bm{u}/|\bm{u}|$. This socalled Skyrmion number does not apply for nonzero disorder potential, when momentum $\bm{p}$ is no longer a good quantum number.

We calculate the topological invariant for nonzero $V$ from the winding number of the reflection matrix $r(\phi)$ in a cylinder geometry,\cite{Braunlich2009,Ful11}
\begin{equation}
{\cal I}=-\frac{1}{2\pi i}\int_{0}^{2\pi}d\phi\,\frac{d}{d\phi}\ln{\rm Det}\,r(\phi),\label{IDetphi}
\end{equation}
where $\Phi=\phi\hbar/e$ is the flux enclosed by the cylinder and $r(\phi)$ is evaluated at $\varepsilon=0$. (We explain in App.\ \ref{app:topdis} how to construct the quasi-energy dependent reflection matrix from the Floquet operator.\cite{Fyo00,Oss02}) Since this is a 2D system, the sizes $M\times M$ for which we can calculate ${\cal I}$ are much smaller than in the 1D reduction used to calculate $D(t)$. 

\begin{figure}[tb]
\centerline{\includegraphics[width=0.9\linewidth]{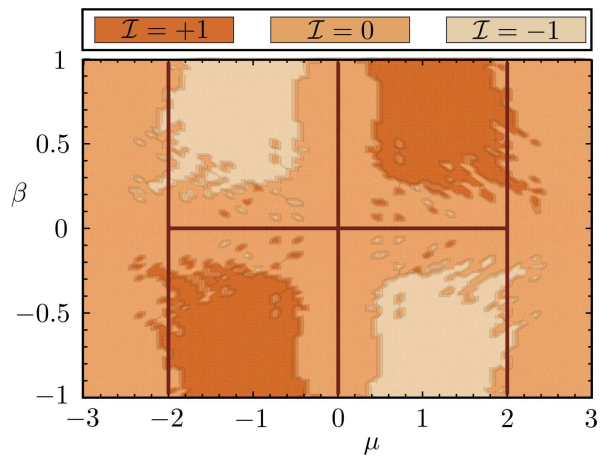}}
\caption{\label{fig:phasediagram}
Phase diagram of the topological invariant ${\cal I}$ in a cylinder of size $40\times 40$, calculated from Eq.\ \eqref{IDetphi} for a single disorder realization. The solid lines are the phase boundaries \eqref{QWZsigmaxy} in the clean system.
}
\end{figure}

The results in Fig.\ \ref{fig:phasediagram} are for $M\times M=40\times 40$. This is data for a single sample ($\bm{q}=0$), at fixed $K=2$ as a function of $\beta,\mu$. The disorder-averaged $\mu$-dependence of ${\cal I}$ is plotted in Fig.\ \ref{fig:peakscan} (dashed curve, for $\beta=0.8$). 

Comparing with the phase boundaries \eqref{QWZsigmaxy} for the clean system ($V\equiv 0$), we see that disorder introduces topologically trivial regions along clean phase boundaries. In the disordered system transitions between two different topologically nontrivial phases (with ${\cal I}=\pm 1$) go via a topologically trivial region (${\cal I}=0$). A similarly disruptive effect of disorder (but with a metallic gapless region replacing the topologically trivial phase) has been observed in computer simulations of the quantum \textit{spin} Hall effect.\cite{Yam11}

We have also calculated the Hall conductance $G_{\rm H}$, which unlike the topological invariant is a directly measurable quantity. The results shown in Fig.\ \ref{fig:peakscan} (data points) were obtained in a single four-terminal sample of dimensions $M\times M=70\times 70$, directly from the scattering matrix expression for the Hall conductance.\cite{Buettiker1986} The Hall plateaus are at the values expected from the topological invariant, $G_{\rm H}\approx{\cal I}\times e^{2}/h$, with deviations from exact quantization due to the relatively small size of the 2D system.
 
\section{Discussion}
\label{conclude}

We have shown how the quantum Hall effect can be modeled in a 1D dynamical system, by using a pair of incommensurate driving frequencies to simulate the effect of a second spatial dimension. This 1D stroboscopic model could become a competitive alternative to the 2D network model for numerical studies of the quantum Hall phase transition,\cite{Cha88} similarly to how the 1D quantum kicked rotator is an alternative to the 3D Anderson model of the metal-insulator transition.\cite{Shepelyansky1983}

Since quantum kicked rotators can be realized using cold atoms,\cite{Cha08,Lemarie2010,Lemarie2009,Moo75} the stroboscopic model might also provide a way to study the quantum Hall effect using atomic matter waves. Cold atoms represent clean and controllable experimental quantum systems, owing to the ability to tune interaction strengths and external potentials.\cite{Bloch2008} Due to the absence of impurities they have long phase coherence times, so their quantum dynamics can be followed over long time scales. These properties make cold atoms ideally suited for the experimental study of quantum phase transitions.

There is a particular need for a new physical system to investigate the quantum Hall phase transition, because currently the theory disagrees with semiconductor experiments on the value of the critical exponent.\cite{Slevin2009} This might be an effect of Coulomb interactions between the electrons in a semiconductor, and a system with controllable interactions could shed light on this question.

For cold atomic gases prepared in a magneto-optical trap a quasi-periodically modulated 1D standing wave, created by two overlapping laser beams, simulates the quasi-periodic driving of the kicked rotator.\cite{Moo75} The momentum distribution is accessible through an absorption measurement, following the release of the atomic gas from the trap.\cite{Bloch2008} Since in the kicked rotator momentum plays the role of coordinate, in this way the diffusion coefficient could be measured and the critical exponent of the metal-insulator transition was obtained from its time dependence.\cite{Cha08,Lemarie2010,Lemarie2009}

To realize the stroboscopic model of the quantum Hall effect, a controllable spin-$1/2$ degree of freedom is needed. Hyperfine levels in alkali or earth alkali atoms can be used for that purpose,\cite{Bloch2008} and arbitrary rotations of this pseudospin have been demonstrated in Cs.\cite{Karski2009} Two overlapping standing waves would produce a purely sinusoidal kicking potential (corresponding to ${\cal T}(u)\equiv 1$), while for flat spin bands higher harmonics are desirable. Fortunately, the topological nature of the phase transition ensures that there is considerable freedom in the choice of the potentials.

Continuing on the path of dimensional reduction proposed here, it is conceivable that the hypothetical 4D quantum Hall effect\cite{Zha01} might also be realized in the laboratory, by adding two more incommensurate driving frequencies.

\acknowledgments

We are indebted to B. B\'{e}ri for helping us to formulate the model, to A. R. Akhmerov, I. C. Fulga, and F. Hassler for showing us how to calculate the topological invariant, and to G. Lemari\'{e} for information on the cold atom experiments. This research was supported by the Dutch Science Foundation NWO/FOM, by an ERC Advanced Investigator Grant, and by the EU network NanoCTM.

\appendix

\section{Tight-binding representation}
\label{sec:HoppingModel}

To gain further insight into the stroboscopic model, we give a tight-binding representation. This will motivate the specific form \eqref{eq:Tarctan} for the function ${\cal T}(u)$ and it will also guide us in the choice \eqref{Vxdef} for the scalar potential $V(\bm{x})$. The derivation follows the same steps as for the quantum kicked rotator.\cite{Fishman1982,Scharf1989}

Including the spin degree of freedom ($s=\pm 1$), we denote the coordinate basis states by $|\bm{m},s\rangle$, such that $x_{i}|\bm{m},s\rangle=m_{i}|\bm{m},s\rangle$ and $\sigma_{z}|\bm{m},s\rangle=s|\bm{m},s\rangle$. The two states $|a_{\pm}\rangle$ defined by
\begin{subequations}
\begin{align}
&{\cal F}_{\bm{q}}  |a_{+}\rangle=e^{-i\varepsilon}|a_+\rangle,\label{alpha+def}\\
&|a_{-}\rangle=e^{iH_{0}}|a_{+}\rangle=e^{i\varepsilon-iV_{\bm{q}}} |a_{+}\rangle \label{app:relat_alpha+to_alpha-}
\end{align}
\end{subequations}
are evaluated just after and just before the kick. [We have abbreviated $V_{\bm{q}}=V(i\partial_{\bm{p}}+\bm{q}).$] Both states evolve with a phase factor $e^{-i\varepsilon}$ in one period $\tau\equiv 1$. The tight-binding representation is expressed in terms of the average
\begin{equation}
|b\rangle=\tfrac{1}{2}\bigl(|a_{+}\rangle+|a_{-}\rangle\bigr).\label{badef}
\end{equation}

The Hermitian operator
\begin{equation}
  W=i\frac{1-e^{iH_{0}}}{1+e^{iH_{0}}}
  =\frac{1}{u}\tan\bigl[\tfrac{1}{2}u{\cal T}(u)\bigr]\bm{u}\cdot\bm{\sigma}\label{AppWdef}
\end{equation}
allows to relate $|b\rangle$ to $|a_{\pm}\rangle$ separately,
\begin{equation}
|b\rangle=\frac{1}{1+iW}|a_{-}\rangle=\frac{1}{1-iW}|a_{+}\rangle.
\label{not_quite_hopping_model}
\end{equation}
Substitution into Eq.\ \eqref{app:relat_alpha+to_alpha-} gives
\begin{subequations}
\label{WbVrelation}
\begin{align}
&(1+iW)|b\rangle=e^{i\varepsilon-iV_{\bm{q}}} (1-iW)|b\rangle\label{WbVrelationa}\\
&\Rightarrow
i\frac{1-e^{i\varepsilon-iV_{\bm{q}}}}{1+e^{i\varepsilon-iV_{\bm{q}} }} |b\rangle =W|b\rangle\label{WbVrelationb}\\
&\Rightarrow
\tan[(\varepsilon-V_{\bm{q}})/2]|b\rangle=W|b\rangle.\label{WbVrelationc}
\end{align}
\end{subequations}

In coordinate representation this gives the tight-binding equations
\begin{equation}
  \sum_{\bm{n}}\sum_{s'} W_{\bm{n}}^{ss'}b_{\bm{m+n}}^{s'}
   +\tan \biggl[\tfrac{1}{2}V(\bm{m}+\bm{q})-\tfrac{1}{2}\varepsilon \biggr]b_{\bm{m}}^{s}=0,
\label{eq:TightBinding}
\end{equation}
with hopping matrix elements
\begin{subequations}
\label{Wmatrix}
\begin{align}
&W_{\bm{n}}^{ss'}=\langle \bm{m},s |W(\bm{p})|\bm{m+n},s'\rangle,\label{Wmatrix1}\\
&W(\bm{p})=\frac{1}{u}\tan\bigl[\tfrac{1}{2}u{\cal T}(u)\bigr]\bm{u}\cdot\bm{\sigma}.\label{Wmatrix2}
\end{align}
\end{subequations}

The tangent term in Eq.\ \eqref{eq:TightBinding} provides a pseudo-random on-site potential, provided that $V(\bm{m}+\bm{q})$ changes from site to site in a way which is incommensurate with the periodicity $\pi$ of the tangent. This is why a simple polynomial $V(\bm{x})$ suffices to produce the localizing effect of a disorder potential.\cite{Fishman1982} 

The role of the Bloch vector $\bm{q}$ is to provide different realizations of the disorder potential, so that a disorder average is effectively an average of $\bm{q}$ over the Brillouin zone. The strength of the disorder potential is varied by varying the parameter $K$, which determines the relative magnitude of kinetic and potential energies: \textit{small} $K$ corresponds to \textit{strong} disorder.

From Eq.\ \eqref{Wmatrix} we see that different choices for ${\cal T}(u)$ lead to different hopping matrix elements, leaving the on-site disorder unaffected. The arctangent form in Eq.\ \eqref{eq:Tarctan} has the simplifying effect of excluding hopping between sites that are not nearest neighbors. For this choice $\tfrac{1}{u}\tan[\tfrac{1}{2}u{\cal T}(u)]\equiv 1$ the hopping matrix elements are given by
\begin{align}
W_{\bm{n}}={}&2\pi K \beta\mu\sigma_z  \delta_{n_1, 0}\delta_{n_2, 0}\nonumber\\ 
&+ \pi K(\pm i\sigma_y-\beta \sigma_z)\delta_{n_1, 0}\delta_{n_2, \pm1}\nonumber\\
&+\pi K  (\pm i\sigma_x-\beta \sigma_z ) \delta_{n_1, \pm 1}\delta_{n_2, 0}.
\label{hopping_mat_elements_for_arctan}
\end{align}

\section{Finite-time scaling}
\label{finitetime}

Following Refs.\ \onlinecite{Slevin2009,Lemarie2009}, we extract the critical exponent $\nu$ from finite-time numerical data by fitting the diffusion coefficient (or, more conveniently, its logarithm) to the scaling law $D(t)=F(t/\xi^{2})$. For finite $t$ the diffusion coefficient is an analytic function of $\mu$. In view of Eq.\ \eqref{eq:CriticalExponent} the variable $(t/\xi^{2})^{1/2\nu}=t^{1/2\nu}u$ is an analytic function of $\mu$, vanishing at $\mu_{c}$. 

We therefore have the two power series
\begin{align}
&\ln D(t) =\ln D_{c}+\sum_{k=1}^{N_{1}} c^{(1)}_k \left( t^{1/2\nu}u \right)^k  + c_{0}t^{-y},\label{fita}\\
&u=\mu-\mu_{c}+\sum_{k=2}^{N_{2}} c^{(2)}_k(\mu-\mu_{c})^k.\label{fitb}
\end{align}
The term $c_{0}t^{-y}$, with $y>0$, accounts for finite-time corrections to single-parameter scaling at the transition point. We put $c_{1}=0$, $c_{2}<0$, to ensure that $D(t)$ as a function of $\mu$ has a maximum at $\mu_{c}$. We then choose integers $N_{1},N_{2}$ and fit the parameters $D_{c},\nu,c_{0},y$ with $c^{(i)}_{k}$ ($i\in\{1,2\}$, $2\leq k\leq N_{i}$) to the $t$ and $\mu$-dependence of $D(t)$, for a given 1D system size $M$.

For the transition around $\mu=0.38$ we took times $t=1.2\cdot 10^4$, $3.3\cdot 10^4$, $8.3\cdot 10^4$, $2.1\cdot 10^5$, $5.2\cdot 10^5$, and $1.3\cdot 10^6 $, with $M=2^{13}=8192$. We averaged over 1000 samples. The quality of the fit is quantified by the chi-square-value per degree of freedom ($\chi^2/\text{ndf}$). We systematically increased $N_{1},N_{2}$ until we arrived at $\chi^2/\text{ndf}\approx 1$. Only the leading order term in Eq.\ \eqref{fitb} was needed for a good fit, so we simply took $u=\mu-\mu_{c}$. The expansion \eqref{fita} did need higher order terms, up to $N_1=6$. We thus obtained $\nu=2.576\pm0.03$ at $\mu_c = 0.387$ with $\chi^2/ndf=1.2$.
A similar analysis was performed for the outer peak in fig.~\ref{fig:peakscan}, resulting in $\nu=2.565\pm0.03$ at $\mu_c = 1.903$ with $\chi^2/ndf=1.01$.

\section{Scattering matrix from Floquet operator}
\label{app:topdis}

To calculate the topological invariant \eqref{IDetphi} we need the reflection matrix $r(\phi)$ in a cylinder geometry, at quasi-energy $\varepsilon=0$ as a function of the flux $\Phi=\phi\hbar/e$ enclosed by the cylinder. This can be obtained from a four-terminal scattering matrix $S$, which relates the wave amplitudes of incoming and outgoing states at the four edges of an $M\times M$ square lattice of sites $(x_{1},x_{2})=(m_{1},m_{2})$, $m_{i}=1,2,\ldots M$. The dimensionality of $S$ is $8M\times 8M$, with the factor of $8$ accounting for four terminals and a twofold spin degree of freedom. The Floquet operator ${\cal F}_{\bm{q}}$ is a $2M^2\times 2M^2$ matrix describing the stroboscopic time evolution of states on the 2D lattice. (We do not make the dimensional reduction to 1D for this calculation.)

When the square is folded into a cylinder, incoming and outgoing states at the left and right edge are related by a phase factor $e^{i\phi}$. This relation can be used to reduce the four-terminal scattering matrix to a two-terminal scattering matrix $\tilde{S}(\phi)$ (which now has dimension $4M\times 4M$). The reflection matrix $r(\phi)$ is a $2M\times 2M$ subblock of $\tilde{S}(\phi)$, relating incoming and outgoing states at the lower edge. We refer to Ref.\ \onlinecite{Ful11} for a computationally efficient way to carry out this general procedure.

What we discuss in this Appendix is how to obtain $S$ from ${\cal F}_{\bm{q}}$. We are faced with the complication that the truncation of the coordinates to a finite range $M$ introduces spurious hopping matrix elements that couple sites near opposite edges (typically within 5--10 sites from the edge). We cannot directly delete these matrix elements from the Floquet matrix without losing unitarity. 

\begin{figure}[tb]
\centerline{\includegraphics[width=0.6\linewidth]{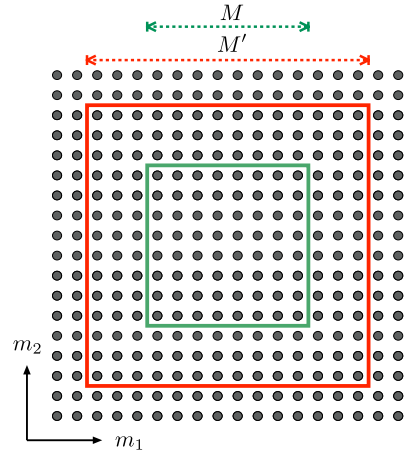}}
\caption{\label{fig:LeadsOnLattice} Truncation of the lattice used to construct a four-terminal scattering matrix, as described in the text.
}
\end{figure}

Our solution (illustrated in Fig.\ \ref{fig:LeadsOnLattice}) is to start from a larger $M'\times M'$ system (red square), with Floquet matrix ${\cal F}'_{\bm q}=e^{-iH'_{0}}e^{-iV'_{\bm{q}}}$. We then go back to the $M\times M$ system (green square) by deleting rows and columns in the coordinate representation of $H'_{0}\mapsto H_{0}$ and $V'_{\bm{q}}\mapsto V_{\bm{q}}$. The resulting Floquet matrix ${\cal F}_{\bm{q}}=e^{-iH_{0}}e^{-iV_{\bm{q}}}$ remains unitary. By choosing $M'$ sufficiently larger than $M$ (typically $M'=M+10$ suffices), we effectively eliminate the spurious hopping matrix elements.

For a four-terminal scattering matrix we introduce absorbing terminals at the four edges of the $M\times M$ lattice. The $8M\times 2M^{2}$ matrix $P$ projects onto these terminals,
\begin{equation}
P^{ss'}_{\bm{mm}'}=\delta_{ss'}\delta_{m_{1}m'_{1}}\delta_{m_{2}m'_{2}}\times\left\{\begin{array}{ll}
1&{\rm if}\;\;m_{1}\in\{1,M\},\\
1&{\rm if}\;\;m_{2}\in\{1,M\},\\
0&{\rm otherwise}.
\end{array}\right.\label{Pdef}
\end{equation}
The $\varepsilon$-dependent scattering matrix $S$ is obtained from the Floquet matrix ${\cal F}_{\bm{q}}$ through the formula\cite{Fyo00,Oss02} 
\begin{equation}
S=P\left[1-e^{i\varepsilon}\mathcal{F}_{\bm{q}}(1-P^T P)\right]^{-1}e^{i\varepsilon}\mathcal{F}_{\bm{q}} P^T,\label{SFrelation}
\end{equation}
where the superscript $T$ indicates the transpose of the matrix. The quasi-energy $\varepsilon$ is set to zero for the calculation of the topological invariant \eqref{eq:Invariantclean}. The integral over $\phi$ is evaluated analytically\cite{Ful11} by contour integration over complex $z=e^{i\phi}$. Results for $M=40$ are shown in Figs.\ \ref{fig:peakscan} and \ref{fig:phasediagram}.

To calculate the Hall conductance $G_{\rm H}$ we directly use the four-terminal scattering matrix $S$, without rolling up the system into a cylinder. The geometry is still that of Fig.\ \ref{fig:LeadsOnLattice}, but the four absorbing terminals are point contacts, covering a single site at the center of each edge. The dimensionality of the scattering matrix, including spin, is thus $8\times 8$. A current $I_{13}$ flows from terminal $1$ to terminal $3$ and the voltage $V_{24}$ is measured between terminals $2$ and $4$ (which draw no current). The Hall conductance $G_{\rm H}=I_{13}/V_{24}$ is obtained from the scattering matrix elements using B\"{u}ttiker's formulas.\cite{Buettiker1986} Results for $M=70$ are shown in Fig.\ \ref{fig:peakscan}.

\end{document}